\documentstyle[prl,aps,amsmath,psfig,amsopn,graphicx]{revtex}

\begin{document}
\draft
\twocolumn
\title{Pattern Formation on the Edge of Chaos: CO Oxidation on 
Pt(110) under Global
Delayed Feedback}
\author{Matthias Bertram, Michael Pollmann, Harm H. Rotermund, Alexander S.
Mikhailov, and Gerhard Ertl}
\address{Fritz-Haber-Institut der Max-Planck-Gesellschaft, Faradayweg 4-6, D-14195 Berlin, Germany}
\date{\today}
\maketitle

\begin{abstract}
Experiments with catalytic CO oxidation on Pt(110) show that chemical 
turbulence in this system can be suppressed by application of 
appropriate global delayed feedbacks. Different spatiotemporal 
patterns, seen near a transition from turbulence to uniform 
oscillations, are investigated. Using a method based on the Hilbert 
transform, spatial distributions of local phase and amplitude in such 
patterns are reconstructed from the experimental data. The observed 
phenomena are reproduced in simulations using a theoretical model 
of the reaction.
\end{abstract}


Control of chaos is one of the central problems in nonlinear dynamics. In
contrast to existing exact methods \cite{Ott90}, a heuristic approach based
on the introduction of delayed feedbacks does not require extensive
real-time computations \cite{Pyragas92}. For extended systems, where
spatially resolved access is difficult, global delayed feedbacks can be
employed. In such methods, information continuously gathered from all
elements is summed up and used to generate a signal which is applied back to
control a common system parameter. Global feedbacks can be employed to
stabilize otherwise unstable trajectories, but also as a tool to produce new
spatiotemporal patterns. Action of global feedbacks on chaotic extended
systems has been experimentally and theoretically investigated for lasers
\cite{Bleich97,Muenkel97}, gas discharges \cite{Pierre96}, semiconductors
\cite{Franceschini99}, populations of electrochemical oscillators \cite{Hudson01}, and surface chemical reactions \cite{Kim01}; it
was also discussed in a general context of the complex Ginzburg-Landau
equation \cite{Batt96,Batt97}. Furthermore, various forms of global feedback
have been successfully applied to control pattern formation in nonchaotic
excitable \cite{Krischer94,Kheowan01,Showalter01} and oscillatory 
\cite{Vanag00,Yang00,Bertram01,Pollmann01} chemical systems.

Nonequilibrium systems on the edge of chaos are capable to generate a broad
variety of complex patterns. To bring a system to a boundary between
deterministic and chaotic dynamics, its parameters may be appropriately
chosen or external forcing may be introduced. However, a practical
implementation of such a predefined control meets serious difficulties
because a system at the edge of chaos is sensitive even to small parameter 
variations. An alternative is provided by using a global
delayed feedback. The advantage of this method is that an acting
force is generated by the system itself and therefore automatically adjusts to the
variations of experimental conditions.

In this Letter, we apply a global delayed feedback to investigate
spatiotemporal pattern formation near a transition to chaos in an
oscillatory surface chemical reaction of catalytic CO oxidation on Pt(110).
Chemical turbulence in this reaction has previously been observed \cite{Jakubith90}. Theoretical analysis has shown the possibility for its existence both under excitable \cite{Bar93} and oscillatory \cite{Falcke94} conditions.
Experiments on suppression of chaos by global 
delayed feedbacks in catalytic CO oxidation have recently been reported \cite{Kim01}. In the present study, the attention is focused on the characterization of complex spatiotemporal patterns resulting from the application of such global
feedbacks. A variety of spatiotemporal structures exhibited upon variation of the feedback intensity and delay time is investigated. 

The catalytic oxidation of carbon monoxide on a platinum(110) single crystal surface follows a relatively simple mechanism
\cite{Imbihl95}. Molecules of CO and oxygen from the gas phase adsorb on the
catalytic surface before the reaction. Adsorbed CO molecules diffuse and
react with immobile adsorbed oxygen atoms to produce carbon dioxide that
immediately leaves the surface. Temporal oscillations of the reaction rate
are due to an adsorbate-driven surface reconstruction in the top substrate
layer. The interplay between reaction and diffusion leads to the development
of spatiotemporal patterns including rotating spiral waves, target patterns,
and turbulence \cite{Jakubith90}.

For the visualization of such patterns, photoemission electron
microscopy (PEEM) is used \cite{Rotermund97}. This method displays the local
work function across a surface area of about 500$\,\mu $m in diameter. In our
experiments, PEEM images were recorded at a rate of 25 per second. The
pressure gauges for O$_{2}$ and CO allowed controlled dosing of reactants
into the reaction chamber. Global delayed feedback was
introduced by making the instantaneous dosing rate of CO molecules dependent
on the properties of imaged concentration patterns. To generate the control
signal $I(t)$, the local PEEM intensity \cite{PEEM} was averaged over the entire observation window. The dosing rate of CO was varied
according to this signal with a delay $\tau _{\mbox{\tiny d}}$. The
variation of the CO partial pressure $p_{\mbox{\tiny CO}}$ in the chamber
followed the temporal modulation of the dosing rate with an additional delay
determined by the residence time of gases in the pumped chamber. Thus, a
controlled feedback was introduced, such that $p_{\mbox{\tiny CO}}(t)=p_{%
\mbox{\tiny 0}}+\mu \,[\,I\,(t-\tau )-I_{\mbox{\tiny 0}}\,]$, where $\tau $
is the effective time delay, the parameter $\mu $ specifies the feedback
intensity, and $p_{\mbox{\tiny 0}}$ and $I_{\mbox{\tiny 0}}$ are the partial
CO pressure and the mean base level of the integral PEEM intensity in
absence of feedback. Gas-phase coupling also leads to
intrinsic variations of partial pressures \cite{Oertzen00}, but these were
significantly smaller than the artificially introduced variation and could be
neglected here.

In the beginning of each experiment, temperature and partial pressures were
chosen in such a way that the reaction was oscillatory and, furthermore, the
unforced pattern represented a state of turbulence where fragments of rotating
spiral waves spontaneously developed and died out at different
locations. After some time the feedback was switched on and its parameters could be
varied. In experiments with a systematic variation of the parameters $%
\mu $ and $\tau $, we have observed that such turbulence could be
suppressed and replaced by stable uniform oscillations for any delay time (delays up to $\tau = 10\,$s were probed) if the feedback intensity $\mu $
was sufficiently high (up to $5\times 10^{-5}\,$mbar, corresponding to CO
partial pressure variations of about 20$\,$\%). Usually, the synchronization
threshold was significantly lower (about 5$\,$\% variations in $p_{%
\mbox{\tiny CO}}$). The period of uniform oscillations was 
affected by the feedback and varied approximately between 3$\,$s and 10$\,$s,
increasing for longer delays and decreasing for
stronger feedbacks.

By fixing the feedback intensity below the transition to uniform
oscillations, various spatiotemporal patterns could be observed. PEEM images of
several typical patterns are displayed in the top row in Fig. 1. Dark (blue) areas are predominantly O covered; bright (red) regions are mainly CO covered. In absence
of feedback, spiral-wave turbulence is observed [Fig.~1(a)]. As $\mu $ is
increased, the feedback becomes effective and global oscillations set on.
Turbulent spiral waves are then replaced by the states of intermittent
turbulence characterized by localized ring-shaped structures or localized spiral-wave fragments on a uniformly oscillating background [Figs.~1(b) and
1(c)]. Such objects repeatedly reproduce until many of them are found, and
again annihilate such that only a few survive. Intermittent turbulence is found independent of the delay time. In addition, for
delays in the interval $0.5\,\mbox{s}\,<\tau <1.0\,$s, 
two-phase clusters [Fig.~1(d)], irregular arrays of cells
[Fig.~1(e)], or oscillatory standing waves [Fig.~1(f)] may develop close to the transition to uniform oscillations.

To characterize the observed patterns, we have employed a variant of the
analytic signal approach \cite{Panter65,Rosenblum98}. This method has allowed us to
transform sequences of typically 250 experimental PEEM images into the time-dependent
spatial distribution of phase and amplitude variables. For the
local PEEM intensity $s({\bf x},t)$ at an observation point ${\bf x}$, its
Hilbert transform $\tilde{s}({\bf x,}t)=\pi ^{-1}\int_{-\infty }^{\infty
}\left( t-t^{\prime }\right) ^{-1}s({\bf x},t^{\prime })dt^{\prime }$ was
computed (this could be easily realized by determining the Fourier transform
of $s$, shifting it by $\pi /2$, and performing the reverse Fourier
transform). This was repeatedly done for all pixel points ${\bf x}$ in an
100 $\times $100 array covering the respective pattern.
Using $s({\bf x},t)$ and its Hilbert transform $\tilde{s}({\bf x,}t)$, a
complex variable $\zeta ({\bf x},t)=s({\bf x},t)+i\,\tilde{s}({\bf x},t)$
was defined. The local oscillation phase $\phi ({\bf x,}t)$ and amplitude $R(%
{\bf x},t)$ were computed as $\phi=\arg\zeta$ and $R=\rho /\rho
_{\rm ref}(\phi )$ where $\rho =\left| \zeta \right| $ and the normalization
to $\rho _{\rm ref}(\phi )$ was introduced to approximately compensate for
deviations from harmonicity in the observed oscillations. To obtain $\rho
_{\rm ref}(\phi )$ the statistical distribution of $\zeta $ for 
all 100$\times $100 pixels and at all 250 time moments was plotted into the complex plane, as illustrated in Fig.~1(g) for a set of spatiotemporal data representing a pattern of spiral-wave turbulence. We determined $%
\rho _{\rm ref}(\phi )$ as a statistical average of $\rho =\left| \zeta
\right| $ inside each of 200 equidistant narrow intervals of the polar angle
$\phi $. Note that the closed curve $\rho =\rho _{\rm ref}(\phi )$ in the
complex plane can be viewed as representing a reference
orbit of the system deduced from the experimental data. 

By applying
this transformation separately to each of the PEEM patterns shown in Fig.~1, spatial
distributions of the phase $\phi$ and amplitude $R$ in
each pattern were constructed. Additionally, Fig. 1
shows a phase portrait of each pattern, obtained by displaying the 
amplitudes and phases for all resolving pixels in polar coordinates.
The phase $\phi$ of a point is represented by the polar angle and the amplitude $R$ is the distance to the coordinate origin.

In spiral-wave turbulence [Fig.~1(a)], the fluctuations of amplitude and
phase are strong, as indicated by the broad-band structure in the phase
portrait, and the amplitude drops down in the spiral cores. 
For intermittent
turbulence [Figs.~1(b) and 1(c)], the amplitude and the phase are almost constant
in the main part of the medium where uniform oscillations take
place. The amplitude is significantly decreased in the ring-shaped objects
[Fig.1(b)] and small localized spirals [Fig. 1(c)], so that they represent
extended amplitude defects. The phase portraits of the intermittent
turbulence show a spot corresponding to the uniform state of the medium
and a tail corresponding to the amplitude defects. In a cluster pattern
[Fig.~1(d)] the medium breaks into two phase states seen as two spots in the
phase portrait. The amplitudes in the two clusters differ
because local oscillations exhibit period-doubling \cite{Kim01}. The
``bridge'' in the phase portrait connecting the two spots corresponds to
the interfaces between the cluster domains; note that the phase varies smoothly and the
amplitude is not significantly reduced at the interface for such cluster
patterns. 
In cellular structures [Fig.~1(e)], small phase modulations 
are observed, while the amplitude remains approximately constant. In
standing waves [Fig.~1(f)], both the phase and the amplitude are
periodically modulated. 

The effects of global delayed feedback on chemical turbulence have been
theoretically investigated using a model \cite{Krischer92,Oertzen00} of the catalytic CO oxidation on Pt(110):
\begin{eqnarray}
\dot{u} &=&k_{1}\,s_{\mbox{\tiny CO}}\,p_{\mbox{\tiny CO}}\,(1-u^{3})-k_{2}%
\,u-k_{3}\,u\,v+D\,\nabla ^{2}u, \\[2mm]
\dot{v} &=&k_{4}\,p_{\mbox{\tiny O$_2$}}\,[\,s_{\mbox{\tiny O,1x1}}w+s_{%
\mbox{\tiny O,1x2}}(1-w)\,](1-u-v)^{2}  \nonumber \\
&&\qquad \qquad \qquad \qquad \qquad \qquad \qquad \qquad -\,k_{3}\,u\,v, \\
\dot{w} &=&k_{5}\,[\,1+\exp \,(\,(u_{0}-u)\,/\,\delta u\,)\,]^{-1}-k_{5}\,w.
\end{eqnarray}
Here, the variables $u$, $v$, and $w$ represent the CO and
oxygen surface coverage and the local fraction of the surface area in the
nonreconstructed state. To account for global delayed
feedback, we assume that the CO partial pressure $p_{\mbox{\tiny CO}}$ in
Eq.~(1) is not constant but varies as $p_{\mbox{\tiny CO}}(t)=p_{{\rm 0}%
}-\mu \,[\,u_{{\rm av}}(t-\tau )-u_{{\rm {ref}}}\,]$, where $u_{{\rm av}}(t)$
is the spatial average of the CO coverage $u({\bf x},t)$ at time $t$, the
parameter $\mu $ specifies the feedback intensity, $\tau $ is the delay
time, $p_{{\rm 0}}$ is the base level of the partial CO pressure, and $u_{{\rm {ref}}}$ is the CO coverage in
the unstable steady state in absence of feedback. The parameters of the
model are chosen in such a way that uniform oscillations are unstable and
amplitude turbulence spontaneously develops without feedback. No-flux
boundary conditions were imposed. Numerical integration of the model
equations yields time-dependent concentration patterns which were further
processed to reconstruct oscillation amplitudes and phases. Because the
simulations already provide two variables, the Hilbert transformation is not
necessary here and a simpler procedure described in Ref.~\cite{Bertram01} was
instead used. 

Figure~2 displays examples of different typical two-dimensional patterns.
The unforced
turbulence [Fig.~2(a)] is characterized by strong amplitude and phase
fluctuations. Stabilization of uniform oscillations by sufficiently strong
feedbacks is found in the model for the delays $\tau >0.1$s. The
intermittent turbulence [Fig.~2(b)] is characterized by irregular cascades
of ''bubbles'' developing into ring-shaped structures on the background of uniform
oscillations. The amplitude is strongly decreased 
inside such localized objects. Stationary two-phase clusters [Fig.~2(c)] are observed under further increase of the feedback
intensity in narrow intervals of the delay time $\tau $. The total area occupied by each of the two cluster domains is equal. Local oscillations
are period-doubled inside the cluster domains. At their interface, the
oscillation amplitude does not vanish, but local oscillations are not period-doubled in the middle of the interface. When $\tau $ is chosen outside of the cluster
intervals (and $\tau >0.1$s), intermittent turbulence is directly
replaced by uniform oscillations upon an increase of $\mu $. 
A hysteresis effect is found in the model: when the feedback intensity $\mu $ is
gradually decreased starting from uniform oscillations, turbulence may set
on only at significantly lower values of $\mu $. Patterns representing regular and irregular oscillatory arrays of cells [Figs.~2(d)
and 2(e)] are then observed in certain intervals of the delay time. These
states are stable with respect to small perturbations, but transform into
amplitude turbulence if stronger perturbations are applied. Both the
phase and the amplitude are modulated in such structures, though the
amplitude variations are weak. 

Thus, our simulations using a realistic model of the CO oxidation reaction
successfully reproduce the principal kinds of patterns seen in the experiments.
Remarkably, the results of our investigations
agree with previous studies \cite{Batt96,Batt97} of global feedbacks in
oscillatory turbulent systems in the framework of the
complex Ginzburg-Landau equation where clusters, oscillatory cellular
arrays, and intermittent turbulence characterized by cascades of
ring-shaped amplitude defects were also found. This
indicates that the observed effects of pattern formation near the edge of
chaos may be typical for a broad class of reaction-diffusion systems. To
characterize the patterns, we have
processed the data to approximately reconstruct amplitude and phase
variables. This representation allows to directly
compare the properties of patterns in systems of different origins and
provides a link to the general studies of turbulence in oscillatory
reaction-diffusion systems \cite{Kuramoto84,Chate96}.

Financial support of the Deutsche Forschungsgemeinschaft under SFB 555
''Complex nonlinear processes'' is acknowledged.


\onecolumn
\begin{figure}[htbp]
  \begin{center}
  \includegraphics[width=17.2cm]{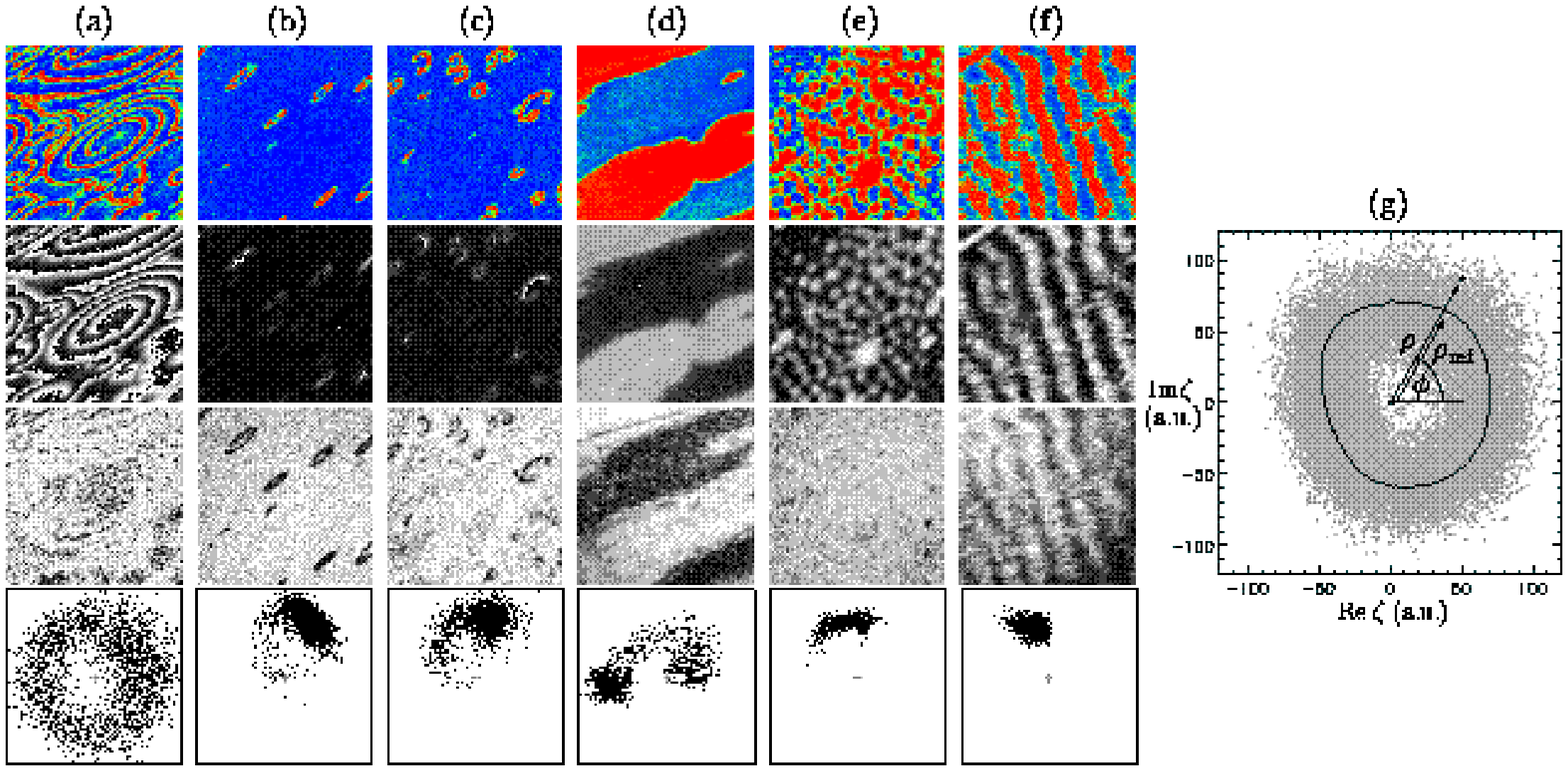} 
  \end{center}
   FIG. 1 (color). (a-f) PEEM images (top), distributions of phase (second row), amplitude (third row), and phase portraits (bottom) for several typical patterns observed in CO oxidation experiments. The values of temperature (K), oxygen partial pressure ($10^{-5}\,$mbar), base CO pressure $p_{\mbox{\tiny 0}}$ ($10^{-5}\,$mbar), feedback intensity $\mu$ ($10^{-5}\,$mbar) and delay time $\tau$ (s) are, respectively: (a) 548. 40.0, 12.3, 0, 0; (b) 540, 40.0, 13.1, 1.7, 0.7; (c) 537, 40.0, 11.4, 3.0, 0.7; (d) 500, 10.0, 3.1, 0.6, 0.8; (e) 535, 40.0, 12.2, 4.0, 0.6; and (f) 505, 10.0, 3.3, 1.6, 0.8. The side length of images is (a,b) 360$\,\mu$m, (c,d) 330$\,\mu$m, (e) 210$\,\mu$m, and (f) 270$\,\mu$m.
(g) Illustration of the transformation to the amplitude $R=\rho /\rho
_{\rm ref}(\phi )$ and phase $\phi=\arg\zeta$ of local oscillations; the reference orbit is indicated. 
  \label{fig1}
\end{figure}

\twocolumn
\begin{figure}[htbp]
  \begin{center}
  \includegraphics[width=8.6cm]{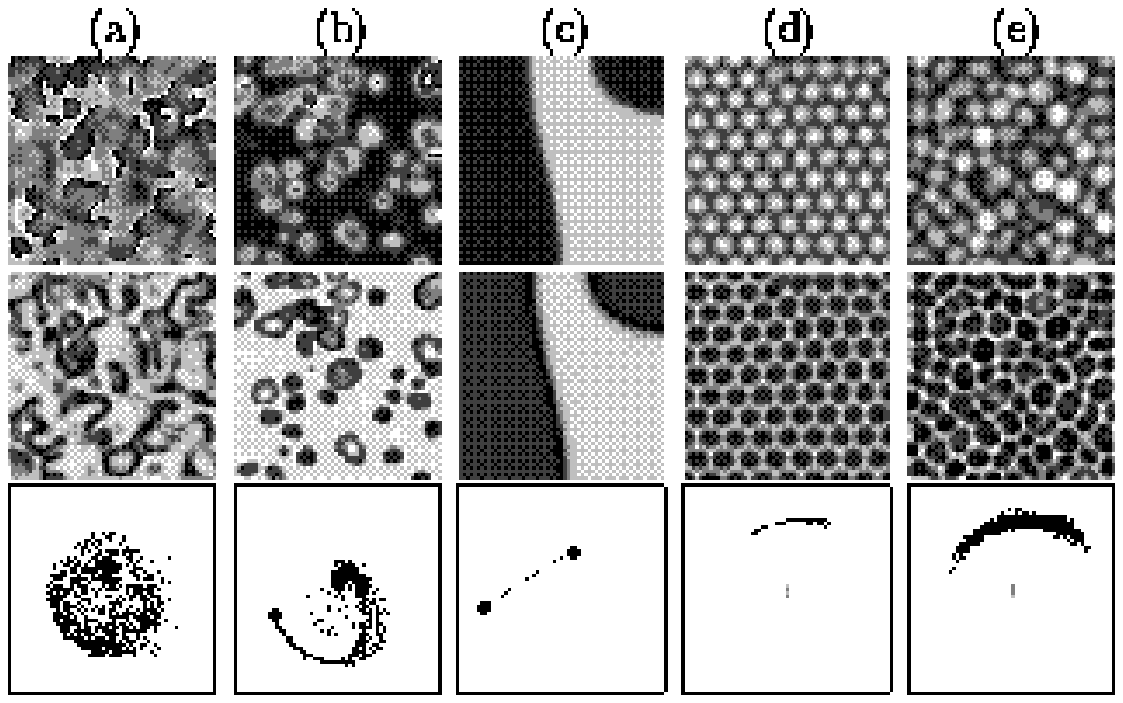} 
  \end{center}
  FIG. 2.
   Distributions of phase (top), amplitude (middle row), and phase portraits (bottom) for several typical simulated patterns. In the phase portraits (b) and (c), bold dots have been added to indicate the uniform states. The model parameters are 
$k_1=3.14 \times 10^5\,$s$^{-1}\,$mbar$^{-1}$, $k_2=10.21\,$s$^{-1}$, $k_3=283.8\,$s$^{-1}$, $k_4=5.86\,$s$^{-1}\,$mbar$^{-1}$, $k_5=1.61\,$s$^{-1}$,
$s_{\mbox{\tiny CO}}=1.0$, $s_{\mbox{\tiny O,1x1}}=0.6$, $s_{\mbox{\tiny O,1x2}}=0.4$, $u_0=0.35$, $\delta u=0.05$, $D=40\,\mu$m$^2\,$s$^{-1}$,   
$p_{\mbox{\tiny O$_2$}}=13.0 \times 10^{-5}\,$mbar, 
$p_{\mbox{\tiny 0}}=4.81 \times 10^{-5}\,$mbar, $u_{\mbox{\tiny ref}}=0.3358$. The values of $\mu$ ($10^{-5}\,$mbar) and $\tau$ (s) are: (a) 0, 0; (b) 0.27, 0.8; (c) 0.40, 0.4; (d) 0.09, 0.3; and (e) 0.06, 0.3. 
The side length is (a, c-e) 400$\,\mu$m and (b) 600$\,\mu$m.
  \label{fig2}
\end{figure}


\begin{thebibliography}{10}

\bibitem{Ott90}
E. Ott, C. Grebogi, and J.~A. Yorke, Phys. Rev. Lett. {\bf 64},  1196  (1990).

\bibitem{Pyragas92}
K. Pyragas, Phys. Lett. A {\bf 170},  421  (1992).

\bibitem{Bleich97}
M.~E. Bleich, D. Hochheiser, J.~V. Moloney, and J.~E.~S. Socolar, Phys. Rev. E
  {\bf 55},  2119  (1997).

\bibitem{Muenkel97}
M. M\"{u}nkel, F. Kaiser, and O. Hess, Phys. Rev. E {\bf 56},  3868  (1997).

\bibitem{Pierre96}
T. Pierre, G. Bonhomme, and A. Atipo, Phys. Rev. Lett. {\bf 76},  2290  (1996).

\bibitem{Franceschini99}
G. Franceschini, S. Bose, and E. Sch\"{o}ll, Phys. Rev. E {\bf 60},  5426
  (1999).

\bibitem{Hudson01}
W. Wang, I.~Z. Kiss, and J.~L. Hudson, Phys. Rev. Lett. {\bf 86},  4954
  (2001).

\bibitem{Kim01}
M. Kim {\it et~al.}, Science {\bf 292},  1357  (2001).

\bibitem{Batt96}
D. Battogtokh and A.~S. Mikhailov, Physica D {\bf 90},  84  (1996).

\bibitem{Batt97}
D. Battogtokh, A. Preusser, and A.~S. Mikhailov, Physica D {\bf 106},  327
  (1997).

\bibitem{Krischer94}
K. Krischer and A. Mikhailov, Phys. Rev. Lett. {\bf 73},  3165  (1994).

\bibitem{Kheowan01}
O.-U. Kheowan {\it et~al.}, Phys. Rev. E {\bf 64},  035201(R)  (2001).

\bibitem{Showalter01}
E. Mihaliuk, T. Sakurai, F. Chirila, and K. Showalter, Discuss. Faraday Soc.
  (to be published).

\bibitem{Vanag00}
V.~K. Vanag {\it et~al.}, Nature {\bf 406},  389  (2000).

\bibitem{Yang00}
L. Yang, M. Dolnik, A.~M. Zhabotinsky, and I.~R. Epstein, Phys. Rev. E {\bf
  62},  6414  (2000).

\bibitem{Bertram01}
M. Bertram and A.~S. Mikhailov, Phys. Rev. E {\bf 63},  066102  (2001).

\bibitem{Pollmann01}
M. Pollmann, M. Bertram, and H.~H. Rotermund, Chem. Phys. Lett. {\bf 346},  123
   (2001).

\bibitem{Jakubith90}
S. Jakubith {\it et~al.}, Phys. Rev. Lett. {\bf 65},  3013  (1990).

\bibitem{Bar93}
M. B\"{a}r and M. Eiswirth, Phys. Rev. E {\bf 48},  R1635  (1993).

\bibitem{Falcke94}
M. Falcke and H. Engel, J. Chem. Phys. {\bf 101},  6255  (1994).

\bibitem{Imbihl95}
R. Imbihl and G. Ertl, Chem. Rev. {\bf 95},  697  (1995).

\bibitem{Rotermund97}
H.~H. Rotermund, Surf. Sci. Rep. {\bf 29},  265  (1997).

\bibitem{PEEM}
The PEEM intensity is defined here in such a way that darker image areas have
  higher intensity.

\bibitem{Oertzen00}
A. von Oertzen, H.~H. Rotermund, A.~S. Mikhailov, and G. Ertl, J. Phys. Chem. B
  {\bf 104},  3155  (2000).

\bibitem{Panter65}
P. Panter, {\em Modulation, Noise, and Spectral Analysis} (McGraw-Hill, New
  York, 1965).

\bibitem{Rosenblum98}
M. Rosenblum and J. Kurths,  in {\em Nonlinear Analysis of Physiological Data},
  edited by H. Kantz, J. Kurths, and G. Mayer-Kress (Springer, Berlin, 1998),
  p.\ 91.

\bibitem{Krischer92}
K. Krischer, M. Eiswirth, and G. Ertl, J. Chem. Phys. {\bf 96},  9161  (1992).

\bibitem{Kuramoto84}
Y. Kuramoto, {\em Chemical Oscillations, Waves, and Turbulence} (Springer,
  Berlin, 1984).

\bibitem{Chate96}
H. Chat\'{e} and P. Manneville, Physica A {\bf 224},  348  (1996).

\end{thebibliography}
\end{document}